\pdfoutput=1
\documentclass[twocolumn,
               showpacs,
               preprintnumbers,
               nofootinbib,
               prd,
               superscriptaddress,
               10pt,
               notitlepage,
               aps]{revtex4-2}
               
\usepackage{graphicx,amssymb,amsmath,amsthm,amsfonts,epsfig,mathtools}

\usepackage[utf8]{inputenc}
\usepackage{graphicx}
\usepackage{dcolumn}
\usepackage{bm}
\usepackage{amsmath}
\usepackage{color}
\usepackage[dvipsnames]{xcolor}
\usepackage{hyperref}
\hypersetup{colorlinks=true, citecolor=MidnightBlue,linkcolor=CornflowerBlue, urlcolor=CornflowerBlue, linktocpage=true}
\usepackage{xfrac}
\usepackage{siunitx}
\usepackage{soul}

\newcommand{\beqn}{\begin{eqnarray}}
\newcommand{\eeqn}{\end{eqnarray}}
\newcommand{\beq}{\begin{equation}}
\newcommand{\eeq}{\end{equation}}

\def\gbar{\bar{g}}
\def\gnn{\bar{g}_{nn}}

\begin{document}

\title{The intrinsic pathology of self-interacting vector fields}

\begin{abstract}
We show that self-interacting vector field theories exhibit unphysical behaviour even when they are not coupled to any external field. This means any theory featuring such vectors is in danger of being unphysical, an alarming prospect for many proposals in cosmology, gravity, high energy physics and beyond. The problem arises when vector fields with healthy configurations naturally reach a point where time evolution is mathematically ill-defined. We develop tools to easily identify this issue, and provide a simple and unifying framework to investigate it. 
\end{abstract}

\author{Andrew Coates}
\email{acoates@ku.edu.tr}
\affiliation{Department of Physics, Ko\c{c} University, \\
Rumelifeneri Yolu, 34450 Sariyer, Istanbul, Turkey}

\author{Fethi M. Ramazano\u{g}lu}
\email{framazanoglu@ku.edu.tr}
\affiliation{Department of Physics, Ko\c{c} University, \\
Rumelifeneri Yolu, 34450 Sariyer, Istanbul, Turkey}

\date{\today}
\maketitle

%%%%%%%%%%%%%%%%%%%%%%%%%%%%%%%%%%%%%%%%
\noindent {\bf \em   Introduction:}
%%%%%%%%%%%%%%%%%%%%%%%%%%%%%%%%%%%%%%%%
Classical electromagnetic waves simply pass through each other when they meet since they obey a linear equation. The picture changes in quantum electrodynamics where two photons can scatter off of each other in principle, they are {\em self-interacting} in this picture \cite{ATLAS:2017fur}. The use of self-interacting vector fields goes beyond this example. They are prevalent in fundamental theories of gravity and cosmology~\cite{Esposito-Farese:2009wbc,DeFelice:2016yws,DeFelice:2016cri,Heisenberg:2017hwb,Kase:2017egk,Ramazanoglu:2017xbl,Annulli:2019fzq,Barton:2021wfj,Minamitsuji:2018kof, Herdeiro:2020jzx,Herdeiro:2021lwl}, and in effective field theories encountered in a wide range of research from astrophysics to plasma physics~\cite{Conlon:2017hhi,Fukuda:2019ewf,dEnterria:2013zqi, Burgess:2020tbq}, including the photon-photon scattering we mentioned~\cite{Heisenberg:1936nmg}. These theories have interesting mathematical structure in their own right~\cite{Loginov:2015rya,Brihaye:2016pld,Brihaye:2017inn,Herdeiro:2021lwl}, and there are systematic efforts to classify all possible self-interacting generalizations of the photon, building on the massive vector theory of Proca~\cite{Proca:1936fbw,Heisenberg:2014rta, Heisenberg:2016eld, Kimura:2016rzw, Allys:2015sht}. In short, self-interacting vector fields can be encountered in all corners of physics. We will, however, show that some of the simplest and most widely encountered forms of vector self-interaction cannot be included in physical theories, hence, many of the ideas we counted above are in need of reevaluation.

The unphysical aspects of self-interacting vector fields arise because their time evolution is not possible beyond a finite duration. Specifically, we show that the field equations that provide the dynamics become unusable, as they no longer define a time evolution. We demonstrate this for vectors that are not coupled to any external fields, which means our results are independent of the context in which the vector is considered, hence they apply to all conceivable cases. These results build on, and widely generalize, a series of studies which first showed that specific self-interacting theories break down near certain astrophysical objects~\cite{Esposito-Farese:2009wbc,Garcia-Saenz:2021uyv,Silva:2021jya,Demirboga:2021nrc,Clough:2022ygm}, and more recently generalized this breakdown to simpler couplings and dynamical cases~\cite{Clough:2022ygm,Mou:2022hqb}. 

A central idea to understand the problem is that the dynamics of the vector field can sometimes be formulated as if governed by a so-called {\em effective metric} that depends on the field itself, even when the gravitational coupling is turned off~\cite{Esposito-Farese:2009wbc,Silva:2021jya,Demirboga:2021nrc,Clough:2022ygm, Clough:2022ygm}. That is, the vector can behave as if in curved spacetime, even when it is not, and this metric can become singular in finite time, at finite vector field values for regular spacetime metrics. 

We show for the first time that the effective metric can be constructed {\em exactly} if spacetime is $1+1$-dimensional, and most likely not in any other case, but surprisingly it still controls the breakdown of time evolution in any dimension. Our approach improves upon earlier approximate methods~\cite{Silva:2021jya,Demirboga:2021nrc,Clough:2022ygm, Clough:2022ygm}, and we also dispel some of the confusion in the literature. We demonstrate that without proper analysis, unphysical coordinate effects can be misidentified as problems in time evolution, or a true breakdown can be overlooked in numerical computations, hence the framework we provide is an essential tool for any future  work on the topic.

These results are highly surprising since they demonstrate that the vector field theories that can exist in nature are tightly constrained, providing a novel appreciation of the Maxwell and Proca theories. We show that heuristic reasoning in field theories, which is commonly based on scalars, can mislead us and mask problems in general, even in the next simplest example of vectors. Furthermore, we show that the analysis of the dynamics of self-interacting vector fields can reveal anomalies that are not apparent in static solutions or a basic counting of the propagating degrees of freedom, hence, it can  be a powerful tool to test a wide variety of theoretical ideas.

Our metric signature is $(-,+,\dots, +)$.

%%%%%%%%%%%%%%%%%%%%%%%%%%%%%%%%%%%%%%%%
\noindent {\bf \em Explicitly hyperbolic formulation of the nonlinear Proca theory:}
%%%%%%%%%%%%%%%%%%%%%%%%%%%%%%%%%%%%%%%%
A simple generalization of the Proca theory, which we dub the {\em nonlinear Proca theory (NPT)}, is given by the Lagrangian
\begin{align}\label{eq:action}
    {\cal L} = -\frac{1}{4} F_{\mu\nu}F^{\mu\nu} - \overbrace{\left( \frac{\mu^2}{2} X^2 + \frac{\lambda \mu^2}{4} \left( X^2\right)^2 \right)}^{V(X^2)} \ ,
\end{align}
where $F_{\mu\nu} = \nabla_\mu X_\nu - \nabla_\nu X_\mu$ and $X^2 =X_\mu X^\mu$ for the real vector field $X_\mu$. 
The corresponding field equation is
\begin{align}\label{eq:eom}
    \nabla_\mu F^{\mu\nu} = 2V' X^\nu = \mu^2 \left(1+\lambda X^2 \right) X^{\nu} \ ,
\end{align}
with $V' = dV/d(X^2)$. We can scale the coordinates and the fields, and without loss of generality set $\mu^2=\pm 1, \lambda=\pm1$ henceforth.\footnote{Despite the notation, $\mu^2$ can have any sign~\cite{Silva:2021jya,Demirboga:2021nrc}.} 

Note that the potential is unbounded from below in some cases. A major point of this study is that the notion of boundedness from below that is central to scalar field theories is insufficient for vectors, as we shall explain. Nevertheless, we still consider $V(X^2)$ to be supplemented by the term $\epsilon (X^2)^4$ for the sake of argument, for some sufficiently small $\epsilon$. One can also physically motivate different parameter signs. For example, $\mu^2=1, \lambda=1$ has a convex self-interaction potential without any intrinsic instabilities, hence is an analog of the nonlinear Klein-Gordon equation. $\mu^2=1, \lambda=-1$ is an effective field theory for the Abelian Higgs mechanism~(e.g \cite{dEnterria:2013zqi, Fukuda:2019ewf,East:2022ppo}).\footnote{The expansion breaks down at $z=0$ but the problems we discuss occur before this point.} $\mu^2=-1, \lambda=-1$ is an analog of the famous Mexican hat potential.

It is not trivial to judge the well-posedness of NPT from Eq.~\eqref{eq:eom} since it is not manifestly hyperbolic, i.e. not in the form of a generalized wave equation. To obtain this form, we first observe that $X_\mu$ obeys the (generalized) Lorenz condition~\cite{Silva:2021jya,Clough:2022ygm}
\begin{align}\label{eq:lorenz}
    \nabla_\nu \nabla_\mu F^{\mu\nu} = 0\ 
    \Rightarrow\ \nabla_\mu \left(z X^\mu \right) = 0 
\end{align}
due to the antisymmetry of $F^{\mu\nu}$, where $z = 2V'/\mu^2=1+\lambda X^2$. 

Using a calculation detailed in Appendix~\ref{sec:gbar}, we show that {\em in $ 1+1$ dimensions} the principal part of Eq.~\eqref{eq:eom} can be rewritten as
\begin{align}\label{eq:gbar_eom}
    \gbar_{\alpha\beta}\nabla^\alpha \nabla^\beta X^\mu + \dots 
    = {\cal M}^{\mu}{}_{\alpha} X^\alpha \ ,
\end{align}
where the ellipses represent single derivative terms, and the effective metric and the mass square tensor are, respectively,\footnote{The overall factor $z$ is optional in the definition of $\gbar_{\mu\nu}$, i.e. our results also hold for $\gbar_{\mu\nu} = g_{\mu\nu} + 2z^{-1}z' X_\mu X_\nu$. }
\begin{align}
    \gbar_{\mu\nu} &= z g_{\mu\nu} + 2z' X_\mu X_\nu \label{eq:g_eff} \\
    {\cal M}^{\mu}{}_{\nu} &= z^2\mu^2 \delta^{\mu}{}_\nu + \textrm{curvature terms}  . \label{eq:M_eff}
\end{align}
We demonstrate in Appendix~\ref{sec:gbar} that, despite some recent approximate computations in $3+1$ dimensions, the above result most likely cannot be generalized beyond $1+1$ dimensions, and we also discuss the exact form of ${\cal M}$. However, the effective metric still determines when the loss of hyperbolicity occurs in any spacetime dimension as detailed in Appendix~\ref{sec:3plus1}.

%%%%%%%%%%%%%%%%%%%%%%%%%%%%%%%%%%%%%%%%
\noindent {\bf \em   The breakdown of time evolution in NPT:}
%%%%%%%%%%%%%%%%%%%%%%%%%%%%%%%%%%%%%%%%
Once it is established that the effective metric governs the dynamics, we immediately see that the time evolution cannot continue to the future of a point where $\gbar_{\mu\nu}$ becomes singular. Hence, our main task is identifying if and when this occurs. 

Our main result is that, starting from problem-free initial data, NPT can naturally evolve to a configuration where the effective metric becomes singular in finite time. Mathematically, this happens when the determinant vanishes
\begin{equation}\label{eq:detg}
    \gbar = g \left(1+\lambda X^2 \right)^d \left(1+3\lambda X^2 \right) = g\ z^d\ z_3 = 0
\end{equation}
where $g = \det(g_{\mu\nu})$, and we used the determinant lemma $\det\left( A + u v^T\right) = (1+ v^T A^{-1} u )\det A$ in $d+1$ dimensions.\footnote{We were informed after the completion of the initial manuscript that a version of this criterion was first used by~\textcite{Esposito-Farese:2009wbc}.} Hence, $\gbar$ vanishes when $z_3=0$, which is encountered earlier than $z=0$ starting from small field amplitudes. Note that the problem is encountered even when $g_{\mu\nu}$ is regular everywhere, and $X^2$ can have either sign, hence, the breakdown is possible for any $\lambda \neq 0$.
We emphasize that a point with $z_3=0$ signifies a physical effect, not a coordinate one. Even though the determinant might vanish due to divergent coordinate transformations in some cases, the physical importance of $z_3=0$ can also be seen in the Ricci scalar of $\gbar_{\mu\nu}$ which can only diverge at a physical singularity. $\bar{R} = F(g_{\mu\nu},\ X_\mu,\ \nabla_\mu X_\nu,\ \nabla_\mu\nabla_\nu X_\rho)/(z\ z_3)$ indeed diverges, since $F$, whose exact form is given in Appendix~\ref{sec:gbar}, is generically nonvanishing at points with $z_3=0$, demonstrating our point. Since $X_\mu$ behaves as if it lives in the spacetime with metric $\gbar_{\mu\nu}$, its time evolution cannot be continued beyond $z_3=0$, the same way any time evolution cannot be continued beyond a spacetime singularity. Lastly, our analysis in Appendix~\ref{sec:3plus1} also identifies $z_3=0$ as the point where hyperbolicity is lost in any dimension, even when the field equations cannot be posed in a manifestly hyperbolic form as in Eq.~\eqref{eq:gbar_eom}. 

We should highlight that the above results only employ the covariant field equation~\eqref{eq:eom} and its necessary implication Eq.~\eqref{eq:lorenz}, hence the loss of well-posedness is not a coordinate effect. The appearance of a curvature singularity additionally signals that there is no formulation of NPT which can evolve beyond this point, see Appendix~\ref{sec:3plus1}.

$z_3=0$ requires the growth of $\lambda X^2$, which can have various causes, e.g. energy transfer to the vector field from an outside source~\cite{Clough:2022ygm}. Since we investigate \emph{intrinsic} pathologies, we do not consider such factors. Rather, we will see that tachyonic instabilities for $\mu^2<0$, or simply the initial ``momentum'' of the fields in terms of nonzero time derivatives suffice. Lastly, note that the growth of the components of $X_\mu$ is not sufficient by itself, since $\lambda X^2$ can stay small or strictly positive, both of which imply $z_3=0$ is not achieved.

$\gbar=0$ is the only form of breakdown in NPT to the best of our knowledge, however there has been another criterion discussed in the recent literature~\cite{Clough:2022ygm,Mou:2022hqb}, which is based on the \emph{$d+1$ decomposition}~\cite{Arnowitt:1962hi,gourgoulhon20123+1}. In this approach we first represent the spacetime as a collection of spatial hypersurfaces in a process called \emph{foliation}, and decompose all tensors into space and time components
\begin{align}\label{eq:adm}
    ds^2 &= -\alpha^2 dt^2 +\gamma_{ij} (dx^i + \beta^i dt)(dx^j + \beta^j dt) \\
    X_\mu &= n_\mu \phi + A_\mu \  ,  \ \phi = -n_\mu X^\mu \  ,  \ A_i=\left(\delta^{\mu}{}_{i} +n^\mu n_i \right)X_\mu \ . \nonumber
\end{align}
The details of this process, some of which can be found in Appendix~\ref{sec:numerics}, is not central to our discussion, aside from the fact that $n^\mu =\alpha^{-1}(1,-\beta^i)$ is a normalized vector field that is orthogonal to the set of spatial hypersurfaces forming our foliation, and defines the slicing of spacetime. $n_\mu \phi$ is orthogonal to the spatial surfaces, and  $A_\mu$ lies on them. Introducing the ``electric field'' $E_i= \left(\delta^{\mu}{}_{i} +n^\mu n_i \right)n^\nu F_{\mu\nu}$, equations~\eqref{eq:eom},~\eqref{eq:lorenz} imply~\cite{Clough:2022ygm}
\begin{equation}\label{eq:eom1p1}
  \begin{aligned}
    \partial_t \phi &= \beta^i D_i \phi -A^i D_i \alpha - \frac{\alpha}{\gnn} z \left(K\phi -D_i A^i \right) \\
     +\frac{2\lambda \alpha}{\gnn} &\left[A^i A^j D_i A_j -\phi \left(E_i A^i - K_{ij} A^i A^j + 2A^i D_i \phi \right) \right] \\
    0&= D_i E^i + \mu^2 z \phi = {\cal C} \ ,
  \end{aligned}
\end{equation}
where
\begin{align}\label{eq:gnn}
    \gnn &= n^\mu n^\nu \gbar_{\mu\nu} = -z+2 \lambda \phi^2 = -z_3 +2\lambda A_iA^i\ .
\end{align}
$D_i$ is the covariant derivative compatible with the induced metric on spatial slices ($\gamma_{ij}$), and $K_{ij}$ and $K$ are the extrinsic curvature and its trace, respectively. ${\cal C}=0$, called the \emph{constraint equation}, is a result of the $\nu=t$ component of Eq.~\eqref{eq:eom}, and does not provide time evolution. However, it has to be satisfied at all times, i.e. on all spatial hypersurfaces.

Recent studies noted that Eq.~\eqref{eq:eom1p1} cannot be solved beyond a point where $\gnn=0$, which was interpreted as a breakdown of time evolution\footnote{This is based on the earliest preprint versions of these studies which may have been updated since.}~\cite{Clough:2022ygm,Mou:2022hqb}. The significance of $\gnn=0$ is that the constraint, ${\cal C}=0$, is a polynomial equation in $\phi$ and the number of roots changes at $\gnn=0$, as $\gnn=\partial{\cal C}/\partial \phi$. This means that $\phi$ will generically be discontinuous at $\gnn=0$, and leads to the more apparent issue that $\partial_t\phi$ diverges.

Before detailing our argument, note that $\gnn = -z_3 +2 \lambda A_i A^i$ implies that for $\lambda>0$, $\gnn=0$ is generically encountered before $\gbar=0$ , and the order is reversed for $\lambda<0$. Thus, for $\lambda<0$ we never encounter $\gnn=0$ during hyperbolic evolution. Thus, we will consider the $\lambda>0$ case in the following discussion.

We believe the issue at $\gnn=0$ to be a \emph{coordinate singularity} which does not imply a physical problem in the time evolution. Namely, $\gnn=0$ arises when one uses a foliation which is not suitable for $\gbar_{\mu\nu}$, possibly because it is adapted to $g_{\mu\nu}$. $\gbar_{\mu\nu}$ controls the dynamics of $X_\mu$, hence the time evolution appears to be problematic for an ill-constructed foliation, similar to coordinate singularities in general relativity~\cite{gourgoulhon20123+1,carroll2004spacetime}. That $\gnn=0$ implies the inability of the solution to satisfy the constraint does not change this fact, since the form of the  constraint equation, hence its root structure, is also foliation-dependent.

Our point can be seen directly in the dependence of $\gnn$ on $\phi=n_\mu X^\mu$, which changes with foliation, unlike $X^2$. Consider a point where $\gnn=0$, $X_\mu = n_\mu \phi + A_\mu$ and $X^2=A_iA^i -\phi^2$ for a foliation defined by the normal vector $n^\mu$. We are free to change our foliation, i.e. choose a new normal vector $\tilde{n}^\mu$, without changing the physics. This provides a new decomposition $X_\mu = \tilde{n}_\mu \tilde{\phi} + \tilde{A}_\mu$.  Then, if $X^2>0$, we can choose $\tilde{n}^\mu$ to be orthogonal to $X^\mu$ so that $\tilde{\phi} = \tilde{n}_\mu X^\mu=0 \Rightarrow X^2 = \tilde{A}_i \tilde{A}^i$. Whereas if $X^2<0$ we can choose $\tilde{n}^\mu$ to be parallel to $X^\mu$ so that $\tilde{\phi} =\textrm{sign} (\phi) \sqrt{\phi^2-A_iA^i} \Rightarrow X^2 = -\tilde{\phi}^2$. In $1+1$ dimensions this can be done globally with some modifications around $X_\mu X^\mu=0$,  but more generally it can at least be performed at the point where $\gnn=0$.  In other words, we can always find a new foliation where $\tilde{A}_i \tilde{A}^i \leq A_iA^i$ (equivalently $\tilde{\phi}^2 \leq \phi^2$), hence $\gbar_{\tilde{n}\tilde{n}} \leq \gnn$, the equality only being possible if $A_i$ vanishes. Thus, in the generic case, the time evolution can be continued in the tilde foliation without issue, thanks to $\gbar_{\tilde{n}\tilde{n}} < 0$, proving our point that $\gnn=0$ is a result of an ill-suited foliation.\footnote{This is not relevant for earlier studies with diagonal effective metrics~\cite{Garcia-Saenz:2021uyv,Silva:2021jya,Demirboga:2021nrc}, for which $\gnn=0$ implies $\gbar=0$.} The exception, $A_i=0$, leads to $\gnn = \gbar_{\tilde{n}\tilde{n}} =-z_3=0$. However, this also implies $\gbar=0$, hence, the time evolution indeed breaks down in this case, not due to $\gnn=0$, but rather due to $\gbar_{\mu\nu}$ becoming singular.

%%%%%%%%%%%%%%%%%%%%%%%%%%%%%%%%%%%%%%%%
\noindent {\bf \em   Numerical results:}
%%%%%%%%%%%%%%%%%%%%%%%%%%%%%%%%%%%%%%%%
We evolved the vector fields of the Lagrangian~\eqref{eq:action} on a $1+1$ dimensional flat spacetime background, $g_{\mu\nu}=\eta_{\mu\nu}$, using a first order formulation as in Eq.~\eqref{eq:eom1p1}. Overall, we confirm that there exist initial data configurations for any value of $(\mu^2, \lambda)$ for which hyperbolicity is lost. Technical details are in Appendix~\ref{sec:numerics}.

\begin{figure}
\begin{center}
\includegraphics[width=.48\textwidth]{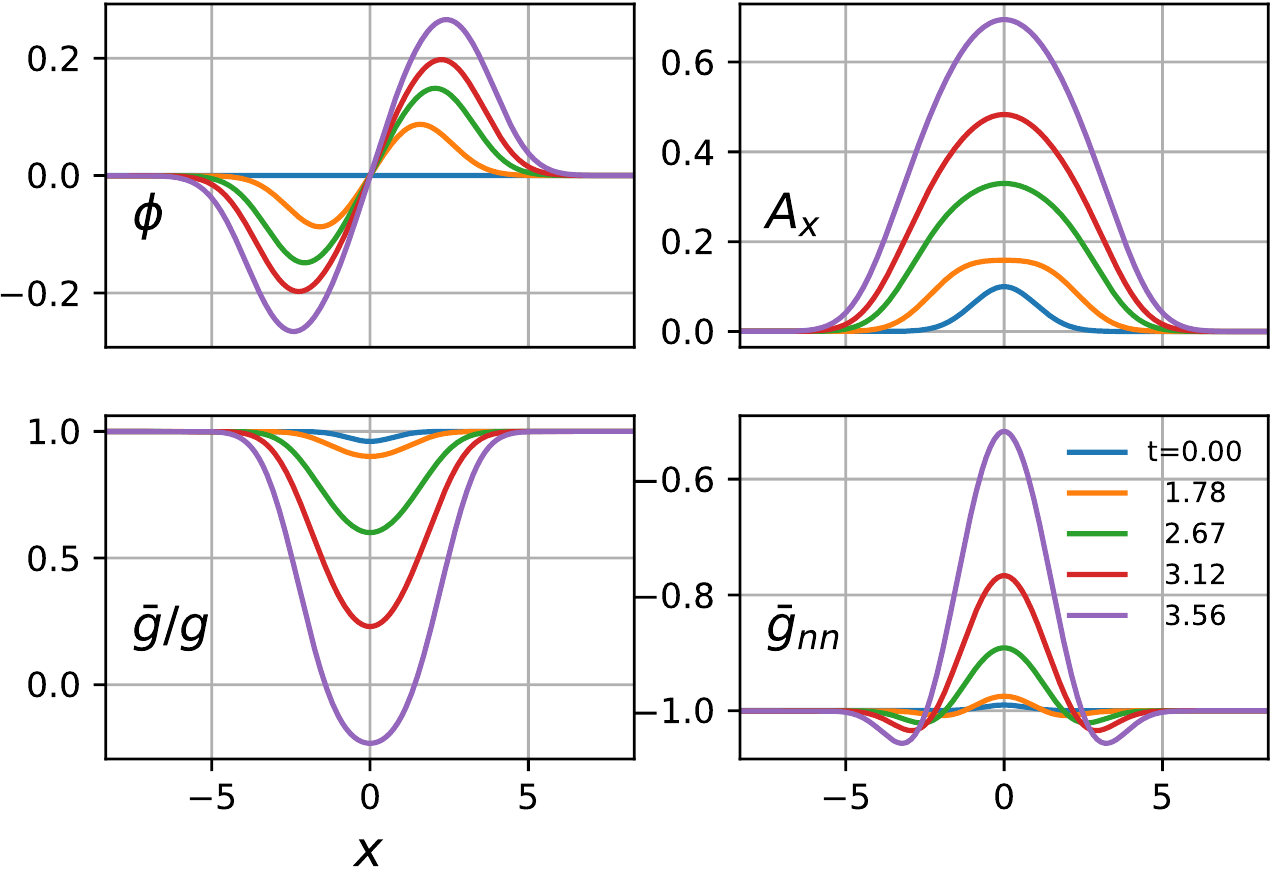}
\end{center}
\caption{Snapshots of $X_\mu$ and $\gbar_{\mu\nu}$ for $\mu^2=-1, \lambda=-1$. The initial growth of the vector is due to a tachyonic instability, which eventually carries the solution to breakdown at $\gbar=0$. The physical meaning of the solution is lost in the region $\gbar>0$, where numerical computation artificially continues due to limited resolution.}
\label{fig:nn}
\end{figure}

\begin{figure}
\begin{center}
\includegraphics[width=.48\textwidth]{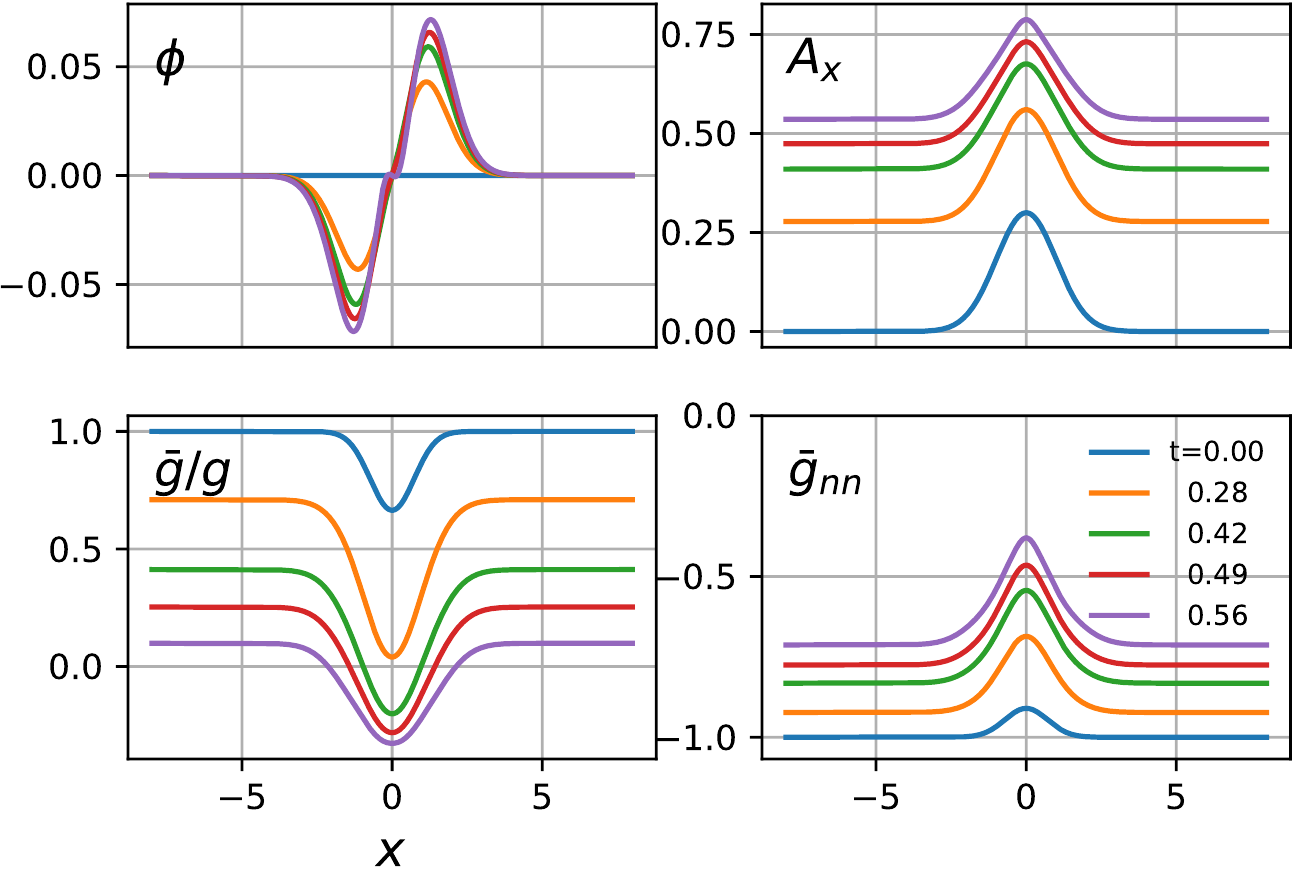}
\end{center}
\caption{Snapshots of $X_\mu$ and $\gbar_{\mu\nu}$ for $\mu^2=1, \lambda=-1$. The initial value of $E_x$ drives $A_x$, and in turn $X^2$, so that hyperbolicity is lost, $\gbar>0$.}
\label{fig:pn}
\end{figure}

Sample evolutions for $\lambda=-1$ can be seen in Figs.~\ref{fig:nn} ($\mu^2=-1$) and~\ref{fig:pn} ($\mu^2=1$), where we encounter $\gbar=0$ without any foliation issues, as expected. The main difference between the cases is that $\mu^2=-1$ breaks down even for arbitrarily low-amplitude initial data due to its tachyonic instability, whereas $\mu^2=1$ requires relatively high initial amplitudes and/or nonzero momentum in the form of $E_x$. Note that the evolution continues beyond $\gbar=0$ as an artifact of the numerics which cannot resolve the problematic fast-growing modes. Hence, these parts of the solutions are not physical (see Appendix~\ref{sec:numerics}).

From a physical perspective, $\mu^2=-1,\lambda=-1$ is a vector analog of the Higgs potential, where the classical ``false vacuum'', $X_\mu=0$, is unstable, but is not dynamically connected to any true vacuum. The effective metric becomes singular well before $X_\mu$ reaches the minimum of $V(X^2)$ at $X^2=1$, at which $\gbar=0$.

The $\lambda=1$ cases require special numerical care since $\gnn=0$ has to be encountered before $\gbar=0$. Even though physical time evolution is not affected by $\gnn=0$, numerical computation fails to continue beyond such a point, hence we cannot investigate the physical breakdown using generic foliations. However, we also saw that, $\gnn=0$ and $\gbar=0$ can be coincident if $A^i=0$ at this point. Therefore, to get as close as possible to $\gbar=0$, we used initial data that satisfies $|A_i|\ll 1$, and $\phi = 1/\sqrt{3\lambda} + \delta \phi$ chosen so that we are already somewhat close to the loss of hyperbolicity. The question is whether the time evolution proceeds towards breakdown starting from this configuration, or away from it.

\begin{figure}
\begin{center}
\includegraphics[width=.48\textwidth]{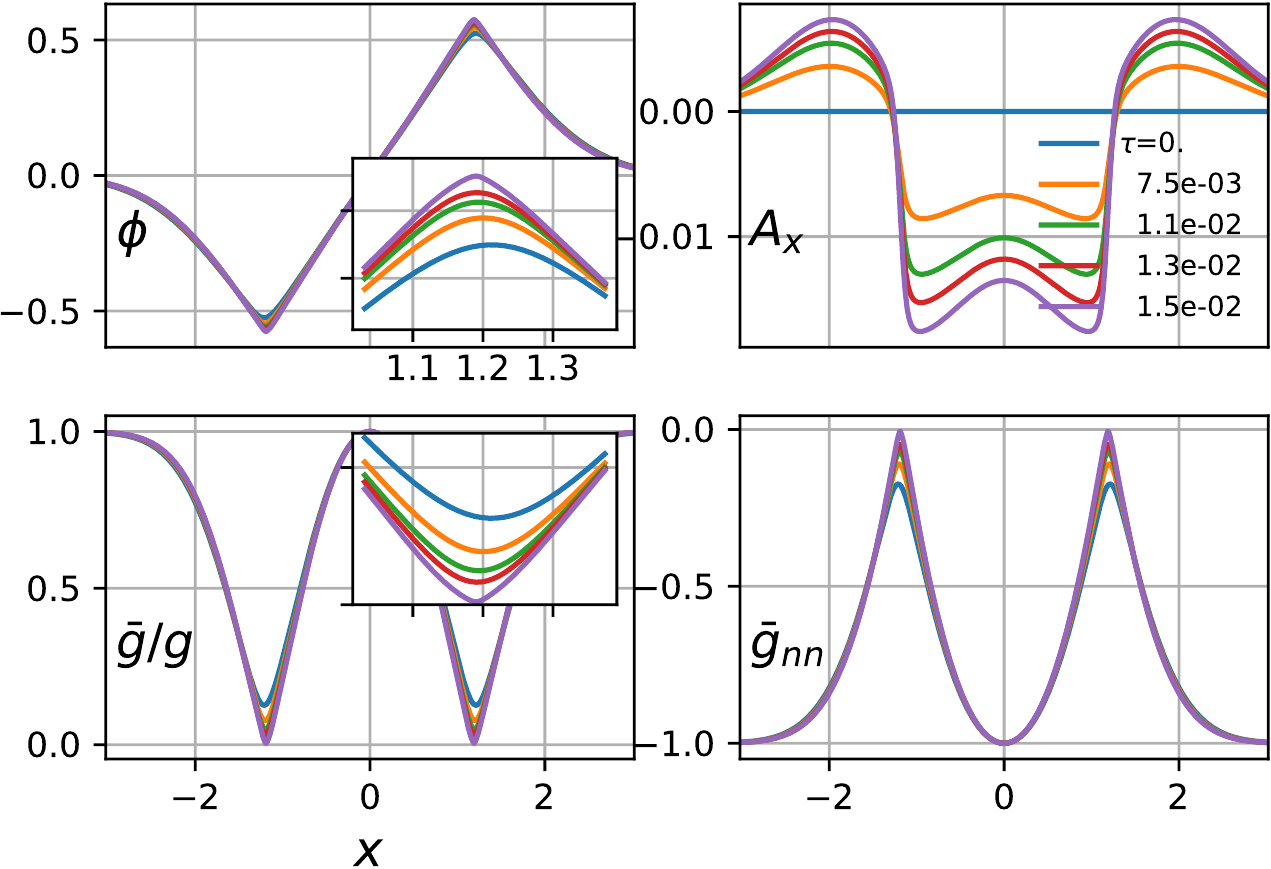}
\end{center}
\caption{Snapshots of $X_\mu$ and $\gbar_{\mu\nu}$ for $\mu^2=1, \lambda=1$. We start with initial data close to breakdown and $A_i=0$. This way, we encounter the coordinate singularity $\gnn=0$ shortly before the true singularity $\gbar=0$, and infer that the solution is indeed evolving towards breakdown.}
\label{fig:pp}
\end{figure}

Analytically, the leading behavior of Eq.~\ref{eq:eom1p1}, $\partial_t \delta\phi= -\left( \sfrac{\alpha K}{9\lambda} \right) (\delta\phi )^{-1} + \dots $, already implies that $\delta\phi$ evolves towards $0$ if $K>0$, which is the case for an appropriate choice of foliation. Thus, we expect the time evolution to break down for $\mu^2=\pm, \lambda=1$. See a sample numerical evolution in Fig.~\ref{fig:pp}.

Lastly, our results can be generalized to any dimensions, e.g. by using our specific initial configurations along one spatial direction and translation symmetry along the rest. Whether more generic initial data can still lead to loss of hyperbolicity in higher dimensions remains to be seen.
%

%%%%%%%%%%%%%%%%%%%%%%%%%%%%%%%%%%%%%%%%
\noindent {\bf \em   Discussion:}
%%%%%%%%%%%%%%%%%%%%%%%%%%%%%%%%%%%%%%%%
The key part of our work was a careful construction of the effective metric, and identifying its singularity as the appropriate criterion for the loss of hyperbolicity. We also revealed the foliation-dependent nature of the commonly used breakdown criterion $\gnn=0$, which can be easily misidentified as a physical breakdown in numerical studies. In essence, the effective metric is generically curved even when the spacetime metric is not. Thus, even in Minkowski spacetime, the usual setting of high energy theories, tools from general relativity are likely required. We explained some of the basic principals for choosing a well-suited foliation for NPT, but future studies will likely require novel approaches.

The problems we revealed can be traced back to the constrained nature of the time evolution and the Lorenz condition, and these do not rely on the specific form of $V(X^2)$, only that it is not linear in $X^2$. Derivative self-interactions also generally lead to generalized Lorenz conditions and constrained evolutions, hence, we expect most, if not all, self-interacting vector field theories to suffer from the same issues.
 
We demonstrate that the intuition gained by studying scalar fields cannot be directly applied to vectors. For example, the $\phi^4$ scalar field theory can be evolved indefinitely for all $(\mu^2,\lambda)$, even if the field amplitude grows without bound. Contrast this with our results, showing that for all $(\mu^2, \lambda)$ the evolution breaks down at finite field values. This is despite the fact that NPT is a member (perhaps the simplest) of the generalized Proca theories, which are explicitly constructed to be ghost free~\cite{Heisenberg:2014rta}. Therefore we suggest that simply counting the degrees of freedom is not sufficient, and our results are essential in investigating the viability of such theories.

All our conclusions about the pathology of NPT considered the theory at face value, i.e. not as an effective approximation to a yet more fundamental theory. Nevertheless, self-interacting vectors, for $\lambda\leq 0$ \cite{Adams:2006sv, deRham:2018qqo}, can appear as such effective fields in some contexts, hence the problems may be resolved in a complete theory. Thus, NPT can still be useful as long as such limitations are taken into account. Exploration of these topics is a lengthy endeavor by itself and an important part of ongoing~\cite{Barausse:2022rvg} and future research.

This study identified the problematic nature of one of the simplest classical field theories, that of a self-interacting vector. We hope our results lead to further research on mathematical constraints on field theories, and the physical implications of such results.

\acknowledgements
We thank Will East for answering our questions about the numerical evolution of nonlinear systems, sharing his ideas about NPT as an effective field theory, and most importantly, for his stimulating questions about the nature of $\gnn=0$. We thank Hector Okada da Silva for pointing out some of the historical literature on the subject. We also thank K{\i}van\c{c} \"Unl\"ut\"urk for reading and commenting on an earlier version of this manuscript. AC acknowledges financial support from the European Commission and T\"{U}B\.{I}TAK under the CO-FUNDED Brain Circulation Scheme 2, Project No. 120C081. FMR was supported by a Young Scientist (BAGEP) Award of Bilim Akademisi of Turkey.

%%%%%%%%%%%%%%%%%%%%%%%%%%%%%%%%%%%%%%%%
\appendix
%%%%%%%%%%%%%%%%%%%%%%%%%%%%%%%%%%%%%%%%
\section{ Derivation of the effective metric} %$\gbar_{\mu\nu}}
\label{sec:gbar}
%%%%%%%%%%%%%%%%%%%%%%%%%%%%%%%%%%%%%%%%
Here, we provide a detailed derivation of the manifestly hyperbolic formulation of NPT field equations~\eqref{eq:gbar_eom} in $1+1$ dimensions. The starting point is the more commonly encountered form of the field equation, Eq~\ref{eq:eom}, which leads to~\cite{Silva:2021jya,Demirboga:2021nrc,Clough:2022ygm}
\begin{align}\label{eq:hyper1}
    0 &= \nabla^\mu F_{\mu\nu} - z \mu^2 X_\nu  \nonumber \\
    &= \nabla^\mu \nabla_\mu X_\nu - \nabla^\mu \nabla_\nu X_\mu -  z \mu^2 X_\nu \nonumber \\
    &= \nabla^\mu \nabla_\mu X_\nu - \nabla_\nu \nabla_\mu X^\mu -R_{\mu\nu}X^\mu - z \mu^2 X_\nu
\end{align}
where we use the definitions of the Riemann and Ricci tensors. The first term on the last line is already the wave operator acting on the vector field, however we need to rewrite the next term, $\nabla_\nu \nabla_\mu X^\mu$, to render the whole equation manifestly hyperbolic. This is typically achieved using the Lorenz condition~\cite{Silva:2021jya,Demirboga:2021nrc,Clough:2022ygm}
\begin{align}
    \nabla_\mu \left(z X^\mu \right) = 0\ \Rightarrow\ 
    \nabla_\mu X^\mu = -\frac{1}{z}X^\mu \nabla_\mu z
\end{align}
Let us insert this into Eq.~\eqref{eq:hyper1}, and only keep track of the second derivative and no derivative terms
\begin{align}\label{eq:hyper2}
    0 &= \nabla^\mu \nabla_\mu X_\nu + \frac{1}{z} X^\mu \nabla_\mu \nabla_\nu z -R_{\mu\nu}X^\mu - z \mu^2 X_\nu +\dots \nonumber \\
    &= \nabla^\mu \nabla_\mu X_\nu + \frac{2z'}{z} X^\mu X^\rho \nabla_\mu \nabla_\nu X_\rho \nonumber\\
    &\phantom{=}-R_{\mu\nu}X^\mu - z \mu^2 X_\nu +\dots \nonumber \\
     &= \nabla^\mu \nabla_\mu X_\nu + \frac{2z'}{z} X^\mu X^\rho \nabla_\mu \nabla_\rho X_\nu + \frac{2z'}{z} X^\mu X^\rho \nabla_\mu F_{\nu\rho} \nonumber\\
    &\phantom{=} -R_{\mu\nu}X^\mu - z \mu^2 X_\nu +\dots \nonumber\\
        &= \left(zg_{\mu\rho}+ 2z' X_\mu X_\rho \right)\nabla^\mu \nabla^\rho X_\nu + 2z' \overbrace{X^\mu X^\rho \nabla_\mu F_{\nu\rho}}^{\Theta_\nu} \nonumber\\
    &\phantom{=} -zR_{\mu\nu}X^\mu - z^2 \mu^2 X_\nu +\dots\ .
\end{align}
Here, we use $\nabla_\mu\nabla_\nu z = \nabla_\nu \nabla_\mu z$ on the first equation, $\nabla_\nu (X_\rho X^\rho) = 2 X^\rho \nabla_\nu X_\rho$ on the second equation, and the definition of $F_{\rho\nu}$ on the third one. We simply group some terms together on the last equation and multiply by an overall factor of $z$, which suggests defining the effective metric
\begin{align}
    \gbar_{\mu\rho} = zg_{\mu\rho} + 2z' X_\mu X_\rho \ .
\end{align}
However, the $\Theta_\nu$ term is also second order, hence contributes to the principal part of the partial differential equation. This means we still need to show that $\Theta_\nu$ can be rewritten in terms of lower order expressions if $\gbar_{\mu\nu}$ is indeed the metric that governs the hyperbolic evolution of $X_\mu$. We were not able to find a general equality that leads to such a result, hence Eq.~\eqref{eq:hyper2} cannot be put into this manifestly hyperbolic form for a generic spacetime to the best of our knowledge. Note that we did not assume anything about the metric or spacetime dimensions until this point, so Eq.~\eqref{eq:hyper2} is valid in all cases nevertheless.

Surprisingly, our specific case of interest, the $1+1$-dimensional spacetime, is an exception to the above null result, that is, we can find a hyperbolic formulation of NPT in this case. For $1+1$-dimensional flat spacetime, $F_{\nu\rho}=\pm\alpha E_x$ when it does not vanish. A straightforward insertion of $\partial_t E_x$ from the evolution equation and $\partial_x E_x$ from the constraint in Eq.~\eqref{eq:eom1p1_long} shows that $\Theta_\nu = 0$. The result can be generalized to any spacetime in $1+1$ dimensions by, for example, using the fact that all such spacetimes are locally conformally flat. Hence, our task for obtaining a hyperbolic equation for $X_\mu$ is completed, for $1+1$ dimensions. $\gbar_{\mu\rho}$ above is indeed the effective metric. 

A central part of this study is identifying the physical singularities of the effective metric, which occur when a curvature scalar diverges. The effective metric is related to the spacetime metric through a so-called \emph{(vector-dependent) disformal transformation} for which the associated Ricci scalars are related as~\cite{Kimura:2016rzw}
\begin{align}\label{eq:ricci_scalar}
    \bar{R} = &\left(\frac{g^{\mu\rho}}{z} - \frac{2 \lambda}{z z_3}X^\mu X^\rho \right) \nonumber \\
    &\times \left(R_{\mu\rho}-2\nabla_{[\mu} B^\nu{}_{\nu]\rho} + 2 B^\alpha{}_{\rho[\mu} B^\nu{}_{\nu]\alpha} \right) 
\end{align}
where $_{[\ ]}$ represents antisymmetrization of indices, and
\begin{align}
    B^\mu{}_{\nu\rho} =  \frac{1}{2}\left(\frac{g^{\mu\sigma}}{z} - \frac{2\lambda}{z z_3}X^\mu X^\sigma \right)
    \left(\nabla_\nu \gbar_{\sigma\rho} + \nabla_\rho \gbar_{\sigma\nu} - \nabla_\sigma \gbar_{\nu\rho} \right)
\end{align}
is the difference in the Christoffel symbols for $g_{\mu\nu}$ and $\gbar_{\mu\nu}$. The factors of $z$ and $z_3$ are apparent in the denominator, and there is no symmetry in the system to cancel these with the terms in the numerator in the generic case. Hence, $\bar{R}$ is divergent and $\gbar_{\mu\nu}$ is singular when $z_3=0$. We also calculated $\bar{R}$ for $g_{\mu\nu}=\eta_{\mu\nu}$ in $1+1$ dimensions, and explicitly checked that it diverges \emph{on shell}, i.e. when the field equations of $X_\mu$ are also taken into account.

\textcite{Clough:2022ygm} recently worked out the steps above to obtain an effective metric for the general spacetime metric in $3+1$ dimensions. They explicitly mention that they neglect a term that is equivalent to $\Theta_\nu$ to obtain the effective metric, which has the same form as ours. We were unable to find any argument to show that $\Theta_\nu$ is negligible compared to the other terms in general, despite our attempts. We should nevertheless emphasize that none of the main results and conclusions in~\textcite{Clough:2022ygm} are directly affected by this discussion, since their numerical formulation, which we adapt, is a $3+1$ decomposition of Eq.~\eqref{eq:eom}, hence does not employ $\gbar_{\mu\nu}$ (which they call $\hat{g}_{\mu\nu}$) in any direct manner. Moreover, we will prove in the next section that $\gbar_{\mu\nu}$ still provides a direct indication of the breakdown of hyperbolicity in any dimension, even when it is not the effective metric in the sense we use the term.

We suggest a simple heuristic argument for the uniqueness of $1+1$ dimensions in the case of hyperbolic formulations for NPT. To reduce the order of $\Theta_\nu$ so that it does not contribute to the principal part of the equations, one needs to substitute out all appearances of derivatives of the electric and magnetic fields. In $1+1$ dimensions, $F$ is a top form and so only contains one degree of freedom, which we call $E_x$. Both $\partial E_x/\partial t$ and $\partial E_x/\partial x $ appear in $\Theta$ but using the two equations of motion these can be directly substituted out, see Eq.~\eqref{eq:eom1p1_long}. In $n$ dimensions there are $n \choose 2$ degrees of freedom in $F$, up to $n$ first derivatives for each and only $n$ equations for reducing the order, and $n\choose 3$ Bianchi identities for relating derivatives, all in all that means there are (up to) $n {n\choose 2} -n - {n\choose 3}$ contributions to the principal part.  For $n>2$ this is larger than $0$. Such a simple counting is likely to overstate the issue as, for example, in $1+1$ dimensions $\Theta$ actually vanishes on shell, which is not immediate from this analysis.

Our discussion so far concentrated on the effective metric, and we overlooked the effective mass squared tensor ${\cal M}^\mu{}_\nu$ in Eq.~\eqref{eq:M_eff}. This tensor is typically considered as the coefficient of the vector term with no derivatives acting on it, ${\cal M}^\mu{}_\nu X_\mu$, and one can consider it to be 
\begin{align}\label{eq:M_exact}
    {\cal M}^\mu{}_{\nu} &= z^2\mu^2 \delta^{\mu}{}_\nu + z R^{\mu}{}_{\nu}
\end{align}
in Eq.~\eqref{eq:hyper2}. This form of the mass square tensor is indeed useful, and can be indicative of the appearance of tachyons as exemplified in many theories~\cite{Silva:2021jya,Demirboga:2021nrc}.

The reason we refrained from providing an exact curved spacetime formula for ${\cal M}$ in Eq.~\eqref{eq:M_eff}, is that, once a coordinate chart is set, the behavior of the differential equation is determined by the partial derivatives rather than the covariant ones, which means covariant derivatives in Eq.~\eqref{eq:hyper2} also contribute to the no-derivative term in the form of the Christoffel symbols. Hence, the behavior is controlled by all such terms, ${\frak M}^\mu{}_\nu$, which is not a tensor in the strict sense any more:
\begin{align}\label{eq:M_partial}
    \mathfrak{M}^\mu{}_{\nu} &= z^2\mu^2 \delta^{\mu}{}_\nu + z R^{\mu}{}_{\nu}  \\
    &\phantom{=} + \gbar_{\tau\rho} g^{\alpha\tau}g^{\beta\rho} \left( -\partial_\alpha \Gamma^{\mu}_{\nu\beta} +\Gamma^{\sigma}_{\alpha\beta}\Gamma^{\mu}_{\sigma\nu} + \Gamma^{\mu}_{\beta\sigma} \Gamma^{\sigma}_{\alpha\nu} \right) \nonumber \\
    &\phantom{=} + \textrm{terms from single covariant derivatives}  \ , \nonumber
\end{align}
where all the terms except the first one are due to curved spacetime or use of curvilinear coordinates. For example, the second line is the contribution of $\gbar_{\tau\rho} \nabla^\tau \nabla^\rho X_\nu$ in Eq.~\eqref{eq:hyper2}. We should further add that, a tachyon is often associated with specific modes in a mode decomposition of the field, which can bring yet more terms that are not present in Eq.~\eqref{eq:M_partial}. The centrifugal barrier term $\ell(\ell+1)/r^2$ that contributes to the effective potential of spherical harmonic modes $(\ell,m)$ in $3+1$ dimensions is a well-known example~\cite{Silva:2021jya,Demirboga:2021nrc}. Finally, we should note that when the effective metric is not known in higher dimensions, the meaning of ${\cal M}$ and $\mathfrak{M}$ are ambiguous since the effect of no-derivative terms are apparent for manifestly hyperbolic form of the equations, but not in general.

%%%%%%%%%%%%%%%%%%%%%%%%%%%%%%%%%%%%%%%%
\section{The role of the effective metric in other dimensions}
\label{sec:3plus1}
%%%%%%%%%%%%%%%%%%%%%%%%%%%%%%%%%%%%%%%%
Here we demonstrate that the effective metric does have an important role in any dimension. We can see this by directly inspecting the principal part of the linearized problem, which should be well posed if the non-linear problem is~\cite{Sarbach:2012pr} (see~\textcite{Esposito-Farese:2009wbc} for an alternative analysis.). Labelling background quantities with subscript $(0)$ and perturbation quantities with a prefix $\delta$ we have,
\begin{align}
    \left(z_{(0)}g^{\mu\nu}_{(0)} +2 z_{(0)}'X^{\mu}_{(0)}X^{\nu}_{(0)}\right)\partial_{\mu}\partial_{\nu}\delta X_\alpha&\nonumber\\
    +2z_{(0)}'X^{\mu}_{(0)}X^{\nu}_{(0)}\partial_{\mu }\delta F_{\alpha\nu}+\mathrm{l.o.t}& = 0\ .
\end{align}
Expanding out $\delta F$ leads to
\begin{align}
   P_\alpha=&  \left(z_{(0)}g^{\mu\nu}_{(0)} +2 z_{(0)}'X^{\mu}_{(0)}X^{\nu}_{(0)}\right)\partial_{\mu}\partial_{\nu}\delta X_\alpha\nonumber\\
     &+2z_{(0)}'X^{\mu}_{(0)}X^{\nu}_{(0)}\left(\partial_{\mu }\partial_\alpha\delta X_\nu - \partial_{\mu }\partial_\nu\delta X_\alpha\right).\nonumber\\
     =&\left(z_{(0)} g^{\mu\nu}_{(0)}\delta^{\beta}_{\alpha}+ 2z'_{(0)}X^\mu_{(0)}X^\beta_{(0)}\delta^{\nu}_{\alpha}\right)\partial_\mu\partial_\nu \delta X_\beta.
\end{align}
From here we will drop the subscripts $(0)$ for convenience. The following notation is based on that of Ref.~\cite{Kovacs:2020ywu}. The key point to note is that the background fields can be taken to be roughly constant on small enough scales, and it is necessary for the ``frozen-coefficients'' problem to be well-posed for the general problem to be well-posed \cite{Sarbach:2012pr}. Then, we can make the replacement $\partial_\mu\partial_\nu \delta X_\beta \to k_\mu k_\nu \chi_\beta$, where $k$ is a wave $4$-vector and $\chi_\beta$ are Fourier amplitudes. Doing so, we can extract the principal symbol $\mathcal{P}(k)$, which one can think of as the inverse of the propagator, from

\begin{align}
    \mathcal{P}(k)^\beta{}_\alpha \chi_\beta &= z g_{\mu\nu}k^\mu k^\nu \chi_\alpha + 2 z' X_\mu k^\mu X^\beta k_\alpha \chi_\beta \nonumber \\
    \Rightarrow
    \mathcal{P}(k)^\beta{}_\alpha &= z g_{\mu\nu}k^\mu k^\nu \delta^\beta_\alpha+2z'(X^\mu k_\mu)X^\beta k_\alpha. 
\end{align}
From a physical perspective, ${\cal P}$ provides the relationship between the wave numbers and frequencies of the modes, i.e. the dispersion relation, through $\det \mathcal{P}(k)=0$.

The symbol is in the form of a matrix plus a bivector, so we can use the matrix determinant lemma as in Eq.~\eqref{eq:detg}. In $d+1$ dimensions we have $\det A = (z g_{\mu\nu}k^{\mu}k^{\nu})^{d+1} $, $A^{-1} = \mathcal{I}/(z g_{\mu\nu} k^\mu k_\nu)$, with $\mathcal{I}$ the $d+1 \times d+1$ identity matrix, finally resulting in
\begin{align}
    \det \mathcal{P}(k)&=(z g_{\mu\nu}k^{\mu}k^{\nu})^d\left(z g_{\alpha\beta} + 2z' X_\alpha X_\beta\right)k^{\alpha}k^\beta\nonumber\\
    &=\left(z g_{\mu\nu}k^{\mu}k^{\nu}\right)^d\left(\gbar_{\alpha\beta}k^\alpha k^\beta\right)\ .
\end{align}
Therefore there are two distinct types of modes, the standard ones, that solve $g_{\mu\nu}k^{\mu}k^{\nu}=0$ and those that solve $\gbar_{\mu\nu}k^{\mu}k^{\nu}=0$, and a separate singular case of $z=0$. The modes governed by $\gbar_{\mu\nu}$ can lead to a loss of hyperbolicity. This means a change of signature of this effective metric directly indicates ill-posedness even when the principal part of the differential equation cannot be written purely in terms of the associated wave operator. Thus, the effective metric can indeed be used to analyze the instability characteristics of time evolution in $3+1$ dimensions, or any other. 

Recall that Eq.~\eqref{eq:detg} and~\eqref{eq:gnn} imply in any dimension
\begin{equation}
    \det \gbar_{\mu\nu}=\det g_{\mu\nu}\ z^d\left(-\gnn +2 \lambda A_i A^i\right)\ ,
\end{equation}
which means that $\gbar=0$ always occurs before $\gnn=0$ for $\lambda<0$, and the order is reversed for $\lambda>0$. As a specific example, it is likely that hyperbolicity is lost due to $\gbar=0$ before $\gnn=0$ is reached in the $\lambda<0$ cases of \textcite{Clough:2022ygm}. Similarly, it is possible that in the $\lambda>0$ case, the time evolution is still well-posed when they encounter $\gnn=0$, and the numerical scheme fails.

We once again emphasize that all results in this section are direct implications of the covariant field equations~\eqref{eq:eom}-\eqref{eq:lorenz}, and do not depend on a specific formulation of the evolution of $X$. Namely, there is a degree of freedom in $X$ that propagates in a spacetime with a naked curvature singularity when $g_{\mu\nu}$ becomes degenerate (see Eq.~\eqref{eq:ricci_scalar}), and one cannot uniquely continue time evolution to the future of such a singularity regardless of what specific formulation is used.

%%%%%%%%%%%%%%%%%%%%%%%%%%%%%%%%%%%%%%%%
\section{Solving the NPT equations}
\label{sec:numerics}
%%%%%%%%%%%%%%%%%%%%%%%%%%%%%%%%%%%%%%%%
\subsection{Field equations and foliation choices}
We use a $d+1$ decomposition of the complete system of field equations on a fixed background metric~\cite{Clough:2022ygm,Zilhao:2015tya}:
\begin{equation}\label{eq:eom1p1_long}
  \begin{aligned}
    {\rm d}_t \phi &= -A^i D_i \alpha - \frac{\alpha}{\gnn} z \left(K\phi -D_i A^i \right) +\frac{\alpha}{\gnn} Z \\
     +\frac{2\lambda \alpha}{\gnn} &\left[A^i A^j D_i A_j -\phi \left(E_i A^i - K_{ij} A^i A^j + 2A^i D_i \phi \right) \right] \\
    {\rm d}_t A_i &=-\phi D_i \alpha - \alpha \left( E_i +D_i \phi \right)\\    
    {\rm d}_t E_i &= D^j \left[ \alpha \left(D_i A_j -D_j A_i \right) \right] \\ 
    &\phantom{=} +\alpha \left(KE_i -2K_{ij}E^j +D_i Z \right) 
    +\mu^2 z \alpha A_i
     \\ 
    {\rm d}_t Z &= -\alpha \left(\kappa  Z - {\cal C}\right) \\
    0&= D_i E^i + \mu^2 z \phi = {\cal C} \\
    z &= 1+\lambda A_i A^i -\lambda \phi^2\ .
  \end{aligned}
\end{equation}
Lapse $\alpha$, shift $\beta$, induced metric $\gamma_{ij}$, extrinsic curvature $K_{ij}$ and its trace $K=\gamma^{ij} K_{ij}$ determine the specific foliation we use for the spacetime whose details can be found in standard references~\cite{gourgoulhon20123+1}. Indices of spatial tensors are raised and lowered with the induced metric $\gamma_{ij}$ and ${\rm d}_t = \partial_t - {\cal L}_\beta$, where ${\cal L}_\beta$ is the Lie derivative along $\beta^i$. We keep our formulation general for $d+1$ dimensions even though our numerical results are for $d=1$. $Z$ is an auxiliary \emph {constraint damping} variable that ensures that the constraint does not grow due to numerical reasons~\cite{Clough:2022ygm,Zilhao:2015tya}.

We discussed the importance of the foliation employed in Eq.~\eqref{eq:eom1p1_long} in order to avoid encountering $\gnn=0$ early in the computation. This means, even if the spacetime metric is flat, one may need nontrivial foliations that are better compatible with the effective metric $\gbar_{\mu\nu}$. $\gnn=0$ is not an issue for $\lambda=-1$ (Figs.~\ref{fig:nn} and~\ref{fig:pn}),  where we used the trivial foliation
\begin{align}\label{eq:slice_trivial}
    ds^2 = -dt^2 + dx^2\ ,
\end{align}
hence, $\alpha=1=\gamma_{xx}$ and $\beta_x=0=K_{xx}$. Spatial tensors have a single component, and we use the index $x=1$ for clarity.

For $\lambda=1$ (Fig.~\ref{fig:pp}), we used
\begin{align}\label{eq:slice_hyperbolic}
    ds^2 = -d\tau^2 + \frac{bx}{1+x^2}d\tau dx+\left(1 -\frac{b^2x^2}{1+x^2} \right) dx^2\ ,
\end{align}
which is a slightly generalized version of the hyperbolic slicing $b^{-2}(\tau-t)^2-x^2=1$~\cite{gourgoulhon20123+1,Zenginoglu:2007jw} with $|b|\leq1$. This means 
\begin{equation}
  \begin{aligned}
    \alpha &= \left( 1 -\frac{b^2x^2}{1+x^2} \right)^{-1/2} \\
    \beta_x &= \frac{bx}{\sqrt{1+x^2}} \\
    \gamma_{xx} &= 1 -\frac{b^2x^2}{1+x^2}=\frac{1}{\gamma^{xx}} \\
    K_{xx} &= b\left(1+x^2 \right)^{-3/2} \left( 1 -\frac{b^2x^2}{1+x^2} \right)^{-1/2}\ .
  \end{aligned}
\end{equation}
We used $b=0.95$ in Fig.~\ref{fig:pp}, which satisfies the $K>0$ condition needed for breakdown. 

Hyperbolic slicing was sufficient to approach the true singularity of $\gbar_{\mu\nu}$ for our current purposes, however, it is not tailored as a foliation compatible with $\gbar_{\mu\nu}$. Exploring such specific coordinate choices will be a priority in future studies.

\subsection{Initial data}
The initial data for the cases that demonstrate the breakdown of time evolution is constructed using the constraint in Eq.~\eqref{eq:eom1p1}
\begin{align}
    \frac{1}{\sqrt{\gamma}}\partial_x \left( \sqrt{\gamma} \gamma^{xx} E_x \right) + \mu^2 \left[1+\lambda \left(\gamma^{xx}A_x^2- \phi^2\right)\right]\phi =0\ , \nonumber
\end{align}
where we used $D_i V^i = \gamma^{-1/2}\ \partial_i (\gamma^{1/2} V^i)$, $\gamma =\det(\gamma_{ij})=\gamma_{xx}$.

In the $\mu^2=\pm 1, \lambda=-1$ examples (Figs.~\ref{fig:nn} and~\ref{fig:pn}), we use the trivial foliation~\eqref{eq:slice_trivial} and 
\begin{align}
    \phi(0,x)&= 0\ ,\  A_x(0,x) = A_A\ e^{-x^2/(2\sigma_A^2)}
\end{align}
The initial data for $E_x$ can be computed trivially as $E_x(0,x)=const$. For $\mu^2=-1$, $\lambda=-1$, the simplest case of $E_x=0$ leads to the breakdown of the time evolution for any $A_A$. This is not the case for $\mu^2=-1$, $\lambda=1$, but choosing a sufficiently negative constant for $E_x(0,x)$ leads to breakdown, see Fig.~\ref{fig:pn}.\footnote{Nominally, this means $E_x$ is not asymptotically vanishing, however this can be changed by a simple argument. Instead of having $E_x(0,t)=const$ everywhere, we can have this condition in a large but finite region around the origin, but let $E_x$ slowly approach zero at further distances after a transition zone where $A_x(0,x)$ is tiny. $\phi$ does not vanish in the transition zone, and has to satisfy the constraint. However, $A_x$ and $D_i E^i$ terms can be made arbitrarily small, and we need to solve 
\begin{align}
     \phi(0,x) \left( 1-\delta_1 +\phi^2(0,x) \right) = \delta_2,
\end{align}
for small $\delta_{1,2}$, which is always possible. The nonzero $\phi$ fields will travel inwards and change our simulation results in principle, but we can choose the transition zone to be far enough that inward moving disturbances cannot reach the actual computation region before the breakdown occurs.} We should also take care not to start with initial data which already has points with the wrong metric signature, which implies $A_A<\frac{1}{\sqrt{3}}$ following Eq.~\ref{eq:detg}. 

For the case of $\mu^2=1$, $\lambda=1$ in Fig.~\ref{fig:pp}, we use the hyperbolic-like foliation~\eqref{eq:slice_hyperbolic} with $b=0.95$ and
\begin{align}
    E_x(0,x) &= A_E\ \sqrt{\gamma}\ e^{-x^2/(2\sigma_E^2)} \ ,\
    A_x(0,x) = 0\ ,
\end{align}
from which $\phi(0,x)$ can be obtained through root finding in the constraint. $A_E$ is chosen such that $\max(|\phi(0,x)|)$ is close to, but less than, $1/\sqrt{3}$. As explained in the main text, this choice ensures that we can explore the region very close to $\gbar=0$ before we encounter $\gnn=0$ and the computation stops. We can choose $A_E$ so that we start from a point arbitrarily close to the breakdown, but the initial data in Fig.~\eqref{fig:pp} is relatively far away to demonstrate that extreme fine tuning is not necessary.

The exact parameters for the sample evolutions are as follows:
\begin{enumerate}
    \item $\mu^2=-1, \lambda=-1$ (Fig.~\ref{fig:nn}): $A_A=0.1$, $E_x(0,x)=0$.
    \item $\mu^2=1, \lambda=-1$ (Fig.~\ref{fig:pn}): $A_A=0.3$, $E_x(0,x)=-1$.
    \item $\mu^2= 1, \lambda=1$ (Fig.~\ref{fig:pp}): $A_E=0.445$.
\end{enumerate}
$\sigma_x=\sigma_A=\sigma_E =1$ in all cases.

%%%%%%%%%%%%%%%%%%%%%%%%%%%%%%%%%%%%%%%%
\subsection{Computational setup}
%%%%%%%%%%%%%%%%%%%%%%%%%%%%%%%%%%%%%%%%
%
\begin{figure}
\begin{center}
\includegraphics[width=.48\textwidth]{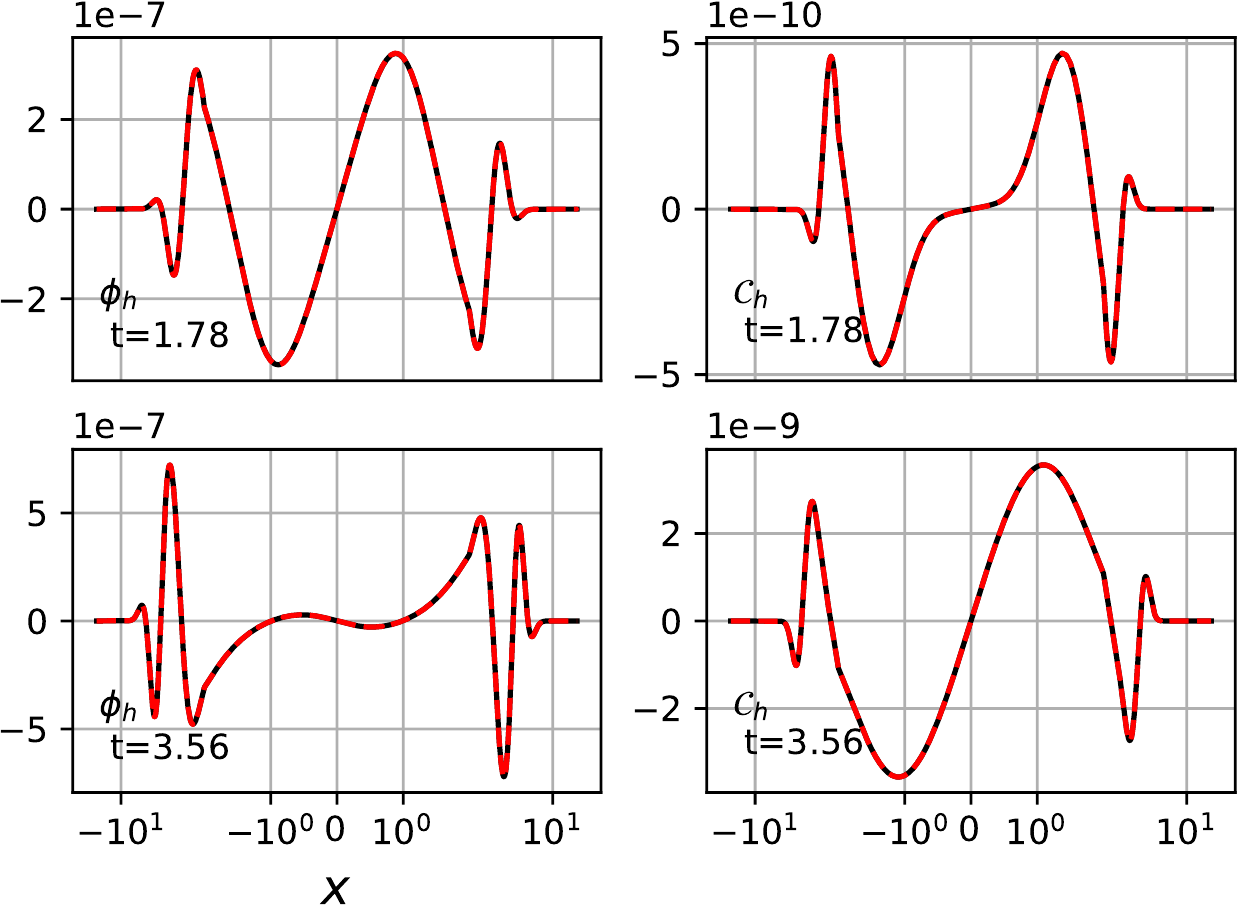}
\end{center}
\caption{4\textsuperscript{th} order convergence of the field components (left column, sample case of $\phi$) and the constraint ${\cal C}$ (right column) for $\mu^2=-1$, $\lambda=-1$ and spatial step size $h=1/64$ in Fig.~\ref{fig:nn}. {\em Left column:}  We check $f_{4h}-f_{2h}$(black) and $16\left(f_{2h}-f_{h}\right)$(red), where $f_h$ is the numerically computed value of the function $f$ ($\phi$ in this case) using step size $h$. These two differences should coincide if the truncation error behaves as ${\cal O}(h^4)$, as is the case for all vector field components. The upper plot shows the convergence at an intermediate stage of the computation, and the lower one at the end. {\em Right column:} Similar to the left, but we directly plot the constraint ${\cal C}_{2h}$(black) and $16{\cal C}_h$(red) rather than the differences between the computations, since ${\cal C}$ is expected to vanish in the continuum limit.}
\label{fig:convergence}
\end{figure}
To solve Eq.~\eqref{eq:eom1p1_long}, we imposed free boundary conditions at a large distance from the region where the initial data was appreciably different from zero, and checked that their effect did not travel into the region around the origin during our computation by considering different box sizes. This is quite wasteful of resources, which is tolerable for $1+1$ dimensions, but future studies in $3+1$ dimensions would likely require compact coordinates, mesh refinement, or both. This and other inefficiencies in our methods can be addressed by the standard tools of numerical relativity when the need arises.

We used 4\textsuperscript{th} order finite differences for spatial derivatives, and the method of lines with the classical 4\textsuperscript{th} order Runge-Kutta method for the time evolution. The simplicity of the one dimensional problem enabled us to use a single grid, and numerical dissipation was not required. This led to robust 4\textsuperscript{th} order convergence of the fields and the constraint in all cases, and a sample of our convergence analysis can be seen in Fig.~\ref{fig:convergence}.

\subsection{Gradient instabilities and convergence pitfalls}
Despite the exemplary convergence in Fig.~\ref{fig:convergence}, one should be wary of the evolution in the cases where hyperbolicity is lost, $\gbar>0$, since it is known that our system of equations do not form a well-posed problem in such regions of spacetime, but the numerics do not immediately crash, and might even seem to converge in low resolution. 

It is instructive to briefly study the nature of the gradient instability in a simple example to realize this. Consider
\begin{align}
    \partial_t^2 u(t,x) = -\partial_x^2 u(t,x) \ ,
\end{align}
where the right hand side has the ``wrong sign.'' A Fourier mode of the form $e^{i(kx-\omega t)}$ has the dispersion relation $\omega = \pm i k$, which means the mode will exponentially grow in time. Even more severely, there is no upper bound to the rate of growth, as higher wave numbers grow faster. This means, even arbitrarily small perturbations can grow to arbitrarily high values in any finite time, and this sensitive dependence on the initial conditions prevent the theory from having any predictive power for the future.

\begin{figure}
\begin{center}
\includegraphics[width=.48\textwidth]{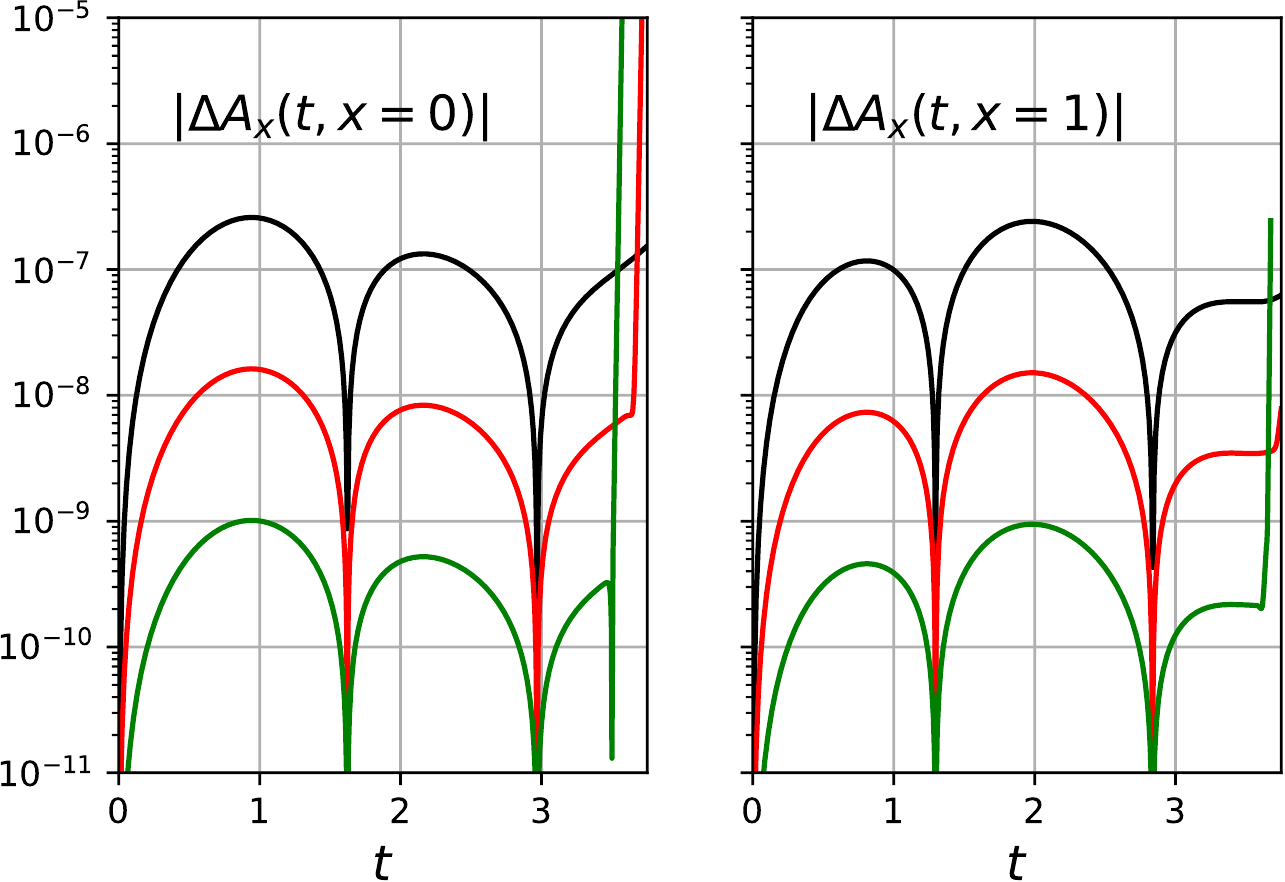}
\end{center}
\caption{Loss of convergence due to the gradient instability for $A_x(t,x=0)$ (left) and $A_x(t,x=1)$ (right) for $\mu^2=-1,\lambda=-1$ (slightly longer evolved version of Fig.~\ref{fig:nn}). We plot the differences between successive resolutions, $|A_{x,8h}(x,t)-A_{x,4h}(x,t)|$ (black), $|A_{x,4h}(x,t)-A_{x,2h}(x,t)|$ (red) and $|A_{x,2h}(x,t)-A_{x,h}(x,t)|$ (green) for $h=1/128$. There is 4\textsuperscript{th} order convergence for most of the simulation, but this is lost some time after the gradient instability (change of metric signature) arises at $t_g=3.3$. More importantly, the convergece is lost closer to $t_g$ for higher resolutions, as explained in the text. The effect is stronger for $x=0$ than $x=1$ since the former is affected longer by the instability (see Fig.~\ref{fig:nn}).}
\label{fig:convergence2}
\end{figure}

The limitations of numerical solutions artificially ameliorate the above picture. The numerical grid is discrete and can only represent wave numbers smaller than $\sim 1/h$, $h$ being the spatial step size. This means there is a bound on the growth rate of the gradient instability, hence one can track the exponential blow up for a limited amount of time, but the solutions always diverge when the computation runs long enough. Perhaps more importantly for numerics, the divergence is increasingly faster for lower $h$ which can accommodate higher wave numbers, see Fig.~\ref{fig:convergence2}. This means the numerical results cannot be trusted in the region where the metric signature changes, even though they may seem to converge for relatively coarse grids.

\bibliography{references_all}

\end{document}